\documentstyle[12pt,a4,oldlfont,epsf]{article}

\begin{document}

\newcommand{\beq}{\begin{eqnarray}}
\newcommand{\ben}{\begin{eqnarray*}}
\newcommand{\eeq}{\end{eqnarray}}
\newcommand{\een}{\end{eqnarray*}}
\newcommand{\s}{\\ \vspace*{-3mm}}
\newcommand{\nn}{\noindent}
\newcommand{\non}{\nonumber}
\newcommand{\ee}{e^+ e^-}
\newcommand{\ff}{\bar{f}f}
\newcommand{\bb}{\bar{b}b}
\newcommand{\aq}{A_{FB}^Q}
\newcommand{\af}{A_{FB}^f}
\newcommand{\ab}{A_{FB}^b}
\newcommand{\ra}{\rightarrow}
\newcommand{\sw}{s_W^2}
\newcommand{\cw}{c_W^2}

\renewcommand{\thefootnote}{\fnsymbol{footnote} }

\nn
\hspace*{11cm} MPI-Ph/94-81 \\
\hspace*{11cm} UdeM-GPP-TH-94-09 \\
\hspace*{11cm} DESY 94--201 \\
\hspace*{11cm} hep-ph/9411386 \\
\hspace*{11cm} October 1994 \s

\vspace*{2cm}

\centerline{\large{\bf A Note on the QCD Corrections to
Forward--Backward }}

\vspace*{5mm}

\centerline{\large{\bf Asymmetries of Heavy--Quark Jets in Z Decays}}

\vspace*{1.5cm}

\centerline{\sc A.~Djouadi$^1$, B. Lampe$^2$ and P.~M.~Zerwas$^3$.}

\vspace*{1cm}

\centerline{$^1$ Groupe de Physique des Particules, Universit\'e de
Montr\'eal,}

\centerline{Case 6128 Suc.~A, H3C 3J7 Montr\'eal PQ, Canada.}

\vspace*{2mm}

\centerline{$^2$ Max Planck Institut f\"ur Physik, Werner Heisenberg Institut,}

\centerline{D--80805 M\"unchen, F.R. Germany.}

\vspace*{2mm}

\centerline{$^3$ Deutsches Elektronen--Synchrotron,
                                                   DESY, Notkestra$\ss$e 85,}

\centerline{D--22603 Hamburg, F.R. Germany.}

\vspace*{2cm}

\begin{center}
\parbox{13cm}
{\begin{center} ABSTRACT \end{center}
\vspace*{0.2cm}

\nn The measurement of the forward-backward asymmetries of heavy quarks
provides one of the most precise determinations of $\sin^2\theta_W$
in $Z$ decays. We discuss in detail the one--loop QCD radiative corrections to
these asymmetries. Results are given for single heavy--quark jet
asymmetries and asymmetries of the thrust axis, as well as for
heavy--quark two--jet final states.}

\end{center}

\renewcommand{\thefootnote}{\arabic{footnote} }
\setcounter{footnote}{0}

\newpage

\subsection*{1.~Introduction}

The heavy quark sector, especially $b$-quark physics, has received
much attention in the high--precision analyses
                                            of the electroweak part of the
Standard Model which are being performed in $Z$ decays at LEP and SLC
[for reviews see Ref.~\cite{R1}].
                           In particular, the measurement of the
forward-backward asymmetry for $b$-quark production $A_{FB}^b$ provides
one
of the most precise determinations of the electroweak mixing angle $\sin^2
\theta_W$ in $Z$ decays \cite{R2}. The precision of the experimental data is
so high that higher--order electroweak and QCD
                         corrections must be incorporated properly. \s

The photonic and the genuine electroweak radiative corrections to quark
forward--backward asymmetries $\aq$ have been discussed thoroughly in
the
literature, Ref.~\cite{R3}. Many features of the radiative
corrections due to strong interactions have also been presented in the past
[4-13]. At the one--loop level the QCD corrections consist of virtual
quark--vertex corrections
and bremsstrahlung corrections where one
of the quarks emits an additional gluon in the final state. This bremsstrahlung
correction will depend in a sensitive way on the quark--jet properties,
i.e. the definition of the jet axis
 and the maximum invariant mass allowed for the final--state
jets. \s

In this paper we will give a systematic and complete overview of the
one--loop QCD corrections to the forward--backward asymmetries for
heavy--quark
production in $e^+e^-$ annihilation near the $Z$ resonance.
                                                We will complete
 the theoretical analysis and also correct some erroneous points
in the literature. We will focus on the
quark--mass effects; with increasing accuracy of the LEP
experiments,
these effects become more and more relevant for the production
of bottom and charm quarks. \s

The results will be presented for two different definitions of the reference
direction: the heavy quark axis which is usually used in analytical
calculations [``single quark--jet asymmetry''],
             and the thrust axis, corresponding to the direction of the most
energetic jet, which is more relevant from the experimental viewpoint.
             We
will also present results for forward--backward asymmetries
of two--jet final states in                                             which
the invariant mass of the jets is less than a fraction $\sqrt{y}$ of    the
total energy. In these two--jet
                       asymmetries, the QCD corrections are much smaller
than for the total (single quark--jet and thrust--axis)
                    asymmetries. Two--jet
asymmetries are therefore very useful to minimize
the errors due to strong
interactions in high--statistics measurements. \s

The paper is organized as follows. In the next section, we will summarize, for
the sake of completeness, the tree--level results for the
forward--backward asymmetries.
In section 3, the QCD corrections will be discussed
for the case in which the reference
directions are either the quark or the
thrust axes,
and the results for the single quark--jet and thrust--axis
                                          asymmetries will be presented. In
section 4, we will define the two--jet forward--backward asymmetries;
results of the QCD corrections will be given for the experimentally
useful range of invariant
masses of the jets in the final state. A short summary in
section 5 will conclude the paper.

\subsection*{2.~Basic set-up}

The differential cross section for the process $\ee \ra \ff$ is a binomial in
$\cos\theta$, where $\theta$ is the angle between $e^-$ and $f$,
\beq
\frac{{\rm d} \sigma}{ {\rm d} \cos\theta}=  \frac{3}{8} (1+\cos^2
\theta) \sigma_U + \frac{3}{4} \sin^2 \theta \sigma_L +\frac{3}{4} \cos\theta
\sigma_F
\eeq
$U(L)$ denote the contributions of unpolarized (longitudinally polarized)
$\gamma/Z$
gauge bosons along the reference axis, and $F$ denotes the difference between
right and left polarizations. The total cross
section is the sum of $\sigma_U$ and $\sigma_L$,
\beq
\sigma_T = \sigma_U+ \sigma_L
\eeq
The forward--backward asymmetry for the full angular range can be
expressed by the ratio
of $\sigma_F$ to $\sigma_T$,
\beq
A_{FB}^f = \frac{3}{4} \frac{\sigma_F}{\sigma_T}
\eeq
In the Born approximation, the parton cross sections $\sigma_T$ and $\sigma_F$
are given by
\beq
\sigma_{T}&=& \beta \textstyle
                    \frac{3-\beta^2}{2} \displaystyle
                                        \sigma_{VV}+\beta^3\sigma_{AA}
   \non \\
    \sigma_{F}&=&\beta^2 \sigma_{VA}
\eeq
with $\beta= \sqrt{1- 4m_f^2/s} =\sqrt{1-\mu^2}$ being
the velocity of the final quarks.
[The quark mass, denoted by $m_f$, will later be defined
 more precisely.] The phase space suppression and the dynamical
$\cal{P}$--wave suppression are entirely taken account of by the
 velocity--dependent coefficients.
    In terms of the electric charge of the
fermion $e_f$ and the
vector and axial-vector couplings of the fermion to the $Z$ boson [$I_f^{3L} =
\pm 1/2$ is the weak isospin, and $\sw=1-\cw \equiv \sin^2\theta_W$],
\ben
v_f = 2I_f^{3L} -4s_W^2e_f \ \  \mbox{and}
                                \hspace{0.5cm} a_f = 2I_f^{3L}
\een
the cross sections, for quark final states, can be written as
\beq
\sigma_{VV} &=& \frac{4 \pi \alpha^2 (s)}{s} e_e^2 e_f^2
+ \frac{G_F \alpha(s)}{\sqrt{2}} \frac{M_Z^2(s-M_Z^2)}{D_Z^2}
e_e e_f v_e v_f + \frac{G_F^2}{32\pi} \frac{M_Z^4s}{D_Z^2}
(v_e^2+a_e^2) v_f^2 \non \\
\sigma_{AA} &=& \frac{G_F^2}{32\pi} \frac{M_Z^4s}{D_Z^2}
(v_e^2+a_e^2) a_f^2  \\
\sigma_{VA} &=& \frac{G_F \alpha(s)}{\sqrt{2}} \frac{M_Z^2(s-M_Z^2)}
{D_Z^2} e_e e_f a_e a_f + \frac{G_F^2}{32\pi}
\frac{M_Z^4s}{D_Z^2} v_e v_f a_e a_f \non
\eeq
where, including the energy--dependent
                            $Z$--boson width, $D_Z^2$ is given by
the Breit--Wigner form
$D_Z^2= (s-M_Z^2)^2 +(s\Gamma_Z/M_Z)^2$.

On top of the $Z$ resonance, the forward--backward asymmetry is dominated by
the
pure $Z$ exchange amplitude so that in the Born approximation
\beq
A_{FB/0}^{f} = \frac{3}{4} \ \frac{2v_e a_e}{v_e^2+a_e^2} \
\frac{2 \beta v_f a_f} {\frac{1}{2}(3-\beta^2) v_f^2 + \beta^2 a_f^2}
\eeq

\subsection*{3.~Asymmetries of single quark jets and thrust axis}

At the one--loop level, the QCD corrections to the forward--backward asymmetry
for quark--pair
production can be divided into three parts: (a) the virtual vertex
correction;
(b) soft gluon bremsstrahlung
                             [the additional gluon in the final state carries
an energy less than $\lambda =2E_g/\sqrt{s} \ll 1$]; and (c) hard gluon
bremsstrahlung [the gluon carries an energy larger than $E_g$].
                                                     The results
for these corrections in $Z$ decays are summarized below. \s

\nn \underline{Virtual corrections and soft bremsstrahlung:} \s

\nn The contribution of the virtual corrections to the forward--backward
asymmetry, $\aq$, is given by
\beq
\delta A_{FB}^Q|_{\rm virtual} = A_{FB/0}^Q \times
                                                    \frac{2\alpha_s}{3\pi}
\frac{2\beta^2a_Q^2-3v_Q^2}{ \frac{3-\beta^2}{2}v_Q^2+\beta^2 a_Q^2}
\frac{1-\beta^2}{\beta} \log \frac{1-\beta}{1+\beta}
\eeq
The infrared singularities
     which are present in both $\sigma_T$ and $\sigma_F$, cancel
in the ratio so that $\delta A_{FB}^Q|_{\rm virtual}$ is free of any
singularities. Note that for quark masses approaching zero, the virtual
corrections vanish. \s

The soft gluon bremsstrahlung in the process $\ee \ra Q\bar{Q}g$, in which
the additional gluon carries an energy less than $\lambda =2E_g/\sqrt{s} \ll
1$, contributes to $\sigma_T$ and $\sigma_F$ by the same amount,
\beq
\sigma_{F,T}^{\rm soft~brems} = \sigma_{F,T} (1 + \delta_{\rm SB})
\eeq
The analytical expression of $\delta_{\rm SB}$ can be found in
Ref.~\cite{R4} for instance. This correction therefore cancels
out from the ratio $\sigma_{F} / \sigma_{T}$, and soft gluon
bremsstrahlung does not contribute to $A_{FB}^Q$ to first order
  in $\alpha_S$,
\beq
\delta A_{FB}^Q|_{\rm soft~brems} = A_{FB/0}^Q \times 0
\eeq
The cancellation of the soft bremsstrahlung correction is physically
plausible since the emission of soft energy quanta cannot affect the
angular direction of the leading heavy quarks, a consequence of
Galilei's law of inertia.

\vspace*{0.5cm}

\nn \underline{Hard bremsstrahlung:} \s

\nn In a first step, when analyzing the semi--inclusive process
\ben
\ee\ra Q\bar{Q}g \ra Q + X
\een
we assume the quark momentum to be measured precisely with the
quark direction to be
                     chosen as the reference axis. With this choice, the
contribution of the hard bremsstrahlung to the forward--backward asymmetry in
$Z$ decays including mass effects, is given by

\beq
\delta A_{FB}^Q|_{\rm hard~brems.} = A_{FB/0}^Q
                                                 \times \frac{\alpha_s}{\pi}
\left[ \frac{R_F}{\beta^2} - \frac{ v_Q^2 R_V + a_Q^2 R_A}{\beta \frac{3-
\beta^2}{2} v_Q^2+\beta^3 a_Q^2} \right]
\eeq

\nn with \cite{R3A,R4}
\beq
R_V &=& \frac{2}{3} \int \int \frac{dx d\bar{x}}{(1-x)(1-\bar{x})} \left[ x^2
+ \bar{x}^2 -\frac{1}{4} \mu^2 \left( 2 \lambda_+ +\lambda_- -\chi p_\perp^2
-\bar{\chi} p^2 -\chi \bar{p}^2 + 2 p\bar{p} \right) \right] \non \\
R_A &=& R_V+ \frac{1}{3} \mu^2 \int \int \frac{dx d\bar{x}}{(1-x)(1-\bar{x})}
\left[ z^2 - 6(1-z) + \frac{3}{2} \mu^2 \lambda_+ \right]\non\\
R_F &=& \frac{2}{3} \int \int \frac{dx d\bar{x}}{(1-x)(1-\bar{x})}  \left[
xp -\bar{x} \bar{p} - \frac{1}{2} \mu^2 z \left( \frac{p}{1-x}- \frac{\bar{p}}{
1-\bar{x}} \right) \right] \non
\eeq
\nn Here, $x$ and $\bar{x}$ denote
                               the quark and antiquark energies in units of the
beam energy, $z=2-x-\bar{x}$ is the scaled gluon energy, $\chi=1/\bar{\chi}=
(1-x)/(1-\bar{x})$ and $\lambda_\pm =\chi + \bar{\chi} \pm 2$.
$p$ and $\bar{p}$ are the momenta of the quark and antiquark relative to the
reference axis and $p_\perp$ is the component of the antiquark momentum
perpendicular to this direction:
\ben
p=\sqrt{x^2-\mu^2} \ \ \ , \ \ \ \bar{p}=\sqrt{\bar{x}^2-\mu^2} \cos\theta_{q
\bar{q}} \ \ \ , \ \ \ p_\perp =\sqrt{\bar{x}^2-\mu^2} \sin\theta_{q \bar{q}}
\een
with
\vspace*{-5mm}
\ben
\cos\theta_{q \bar{q}} = \frac{ 2(1-x-\bar{x})+x \bar{x} +\mu^2}{ \sqrt{
x^2-\mu^2}\ \sqrt{\bar{x}^2-\mu^2} }
\een

\nn Using the more convenient variables
                                    $z$ and $w=x -\bar{x}$ the phase space
integration has to be performed over
$ |w|< z \left[ 1-\mu^2/(1-z) \right]^{1/2}$ and $ 2 \lambda < z < 1-
\mu^2 $.   \s

The first contribution in Eq.~(13) accounts for the QCD corrections
of the genuine vector--axial vector interference term while
the second part is due to the corrections to the total cross section. \s

However, due to fragmentation effects, cascade decays and detector
imperfections, the quark direction cannot always be determined very
well. In this realistic situation, the thrust axis lends itself to
the reference axis of
the forward--backward asymmetry. While the thrust axis coincides with
the quark/antiquark line in quark--pair production to lowest order,
this connection is destroyed
                           by hard--gluon emission. In fact,
with a small probability the gluon carries even more energy than
the quarks/antiquarks in three--jet events.\s

The experimental procedure of choosing the thrust axis and
its orientation introduced in Ref.~\cite{R11}, coincides with the
definition of the reference axis in this theoretical analysis.
The orientation of the thrust axis $\vec{T}$ is defined in
such a way that $\vec{T} \cdot \vec{p}$ is positive
where $\vec{p}$ is the 3-momentum of the $b$ quark (Fig.~1).\s

In the three different regions of the Dalitz plot we have for
 the momenta $p, \bar{p}$ and $p_\perp$: \s

$\bullet$ $x > \bar{x},z$ [reference axis parallel to quark line]
\ben
p=\sqrt{x^2-\mu^2} \ \ \ , \ \ \ \bar{p}=\sqrt{\bar{x}^2-\mu^2} \cos\theta_{q
\bar{q}} \ \ \ , \ \ \ p_\perp =\sqrt{\bar{x}^2-\mu^2} \sin\theta_{q \bar{q}}
\een

$\bullet$ $\bar{x} > x,z$ [reference axis anti--parallel to
                                                         antiquark line]
\ben
\bar{p}=\sqrt{\bar{x}^2-\mu^2} \ \ \ , \ \ \ p=\sqrt{x^2-\mu^2} \cos\theta_{g
\bar{q}} \ \ \ , \ \ \ p_\perp =0
\een

$\bullet$ $z > x,\bar{x}$ [reference axis anti--parallel to gluon line]
\ben
p=\sqrt{x^2-\mu^2} \cos \theta_{qg} \ \ \ , \ \ \ \bar{p}=\sqrt{\bar{x}^2-
\mu^2} \cos\theta_{\bar{q} g} \ \ \ , \ \ \ p_\perp =\sqrt{\bar{x}^2-\mu^2}
\sin\theta_{\bar{q} g}
\een
with
\ben
\cos\theta_{qg} = \frac{ 2(1-x-z)+x z }{ z \sqrt{x^2-\mu^2} } \ \ \
\mbox{and}
\ \ \ \ \cos\theta_{\bar{q}g} = \frac{ 2(1-\bar{x}-z)+ \bar{x} z }{ z
\sqrt{\bar{x}^2-\mu^2} }
\een

\nn In addition,
$R_F$ has to be multiplied by $(-1)$ when the reference axis coincides
 with the
antiquark or gluon axis [change from $\cos \theta$ to $-\cos \theta$]; in
$R_V$, the term $\chi p_\perp^2$ has to be replaced by $\lambda_+ p_\perp^2$
when the gluon axis is chosen as
     the reference axis. Note that according to our
prescription for the orientation of the thrust axis, the first two
cases give identical results. Events in which the gluon
3-momentum coincides with the thrust axis, dilute the
forward--backward asymmetry since $\int\sigma_F$ integrated
over the proper part of the Dalitz plot must vanish for this class
of events. \s

The sum of the virtual soft corrections and the hard gluon
                                     bremsstrahlung, integrated
over the entire allowed range, will change the tree--level forward--
backward
asymmetry of single quark--jets to
\beq
A_{FB}^Q = A_{FB/0}^Q \left[ 1 - \frac{\alpha_s}{\pi} C_F^Q \right]
\eeq

The coefficients $C_F^Q$ [which are free of any singularities and
 independent of the gluon energy cut separating the soft from
the hard
bremsstrahlung] are shown in Table 1 for the production of quarks with
nearly zero masses, $c$-quarks and $b$-quarks.
Two reference directions have been chosen: the quark axis               and the
thrust axis. The numbers to the left of the bars correspond to the
pole masses of the quarks, $m_c = 1.5$ GeV and $m_b = 4.5$ GeV; the
numbers to the right to the choice
\cite{R11A} $m_c = 0.7$ GeV and $m_b = 3$ GeV for the
$c,b$ mass parameters, the mass values in the $\overline{MS}$
scheme at the scale $M_Z$. The difference between the two predictions
reflects the uncertainties due to the corrections of
next--to--leading order which have not been included in this analysis.
These uncertainties in the prediction of the $FB$ asymmetries, of order
$10^{-4}$, are {\it much} smaller than the anticipated ultimate
 error
of the LEP measurements.
             \s

\begin{table}[hbt]
\begin{center}
\begin{tabular}{|c||c|c|c|} \hline
& & & \\
\ \ Ref. Axis \ \ & \ \ \ \ $C_F^{m=0} $ \ \ \ \ &   \ \ \ \
$C_F^{c}$ \ \ \ \ & \ \ \ \ $C_F^b$ \ \ \ \ \\
     &       &       &       \\ \hline
     &       &       &       \\
Quark & 1.00 & 0.93 $|$ 0.96 & 0.80 $|$ 0.86 \\
     &       &       &        \\ \hline
     &       &       &       \\
Thrust & 0.89 & 0.86 $|$ 0.88 & 0.77 $|$ 0.81 \\
     &       &       &        \\ \hline
\end{tabular}
\end{center}
\caption{\it Coefficients $C_F^Q$ of
$(-\alpha_s/\pi)$ in the single
quark--jet forward-backward asymmetry with the reference axis being
the quark axis, and for the thrust axis; left: pole quark masses,
right: $\overline{MS}$ masses at the scale $M_Z$.}
\end{table}

One first notices that the coefficient $C_F^Q$ for a given quark species
                    is smaller
when the reference axis is defined as the thrust axis. However, while the
difference is about $\sim 10\%$ for light quarks it decreases to $\sim 5\%$
in the case of $b$ quarks with $m_b=4.5$ GeV.
The numerical results in the case where the
quark axis is chosen as the reference axis, agree with those given in
Refs.~[5--7], and the result in the massless case for the thrust axis
being the reference axis agrees with  Ref.~\cite{R9} if the
pole masses are chosen.\s

The prediction $(- \alpha_s/\pi)$ for the QCD correction of the $FB$
asymmetry in the massless case
has  been noticed first in Ref.~\cite{R4}. The simple coefficient
$(-1)$ is a consequence of the fact that the QCD corrections to
$\sigma_F$ vanish so that the $FB$ asymmetry is only
affected by the well--known
                QCD correction $1 + \alpha_s/\pi$ to the
total cross section in the denominator.\s

Note also that in the case where the quark axis is chosen as a reference, the
coefficient $C_F^Q$ is $\simeq 20\%$
 smaller for $b$--quarks with $m_b = 4.5$ GeV than for massless quarks
[as anticipated from Galilei's law of inertia].
Therefore, although $\mu_b=2m_b/M_Z \sim 0.1$, quark mass effects are vey
important for the QCD corrections. For this choice of the reference axis, the
coefficient $C_F^Q$ has first been calculated numerically
in Ref.~\cite{R4} for arbitrary quark masses, later it has been
derived analytically \cite{R7}. A very good approximation can
be obtained by making an expansion\footnote{Due to a small
error in the analytic
expressions of the integrated parton cross section of
Ref.~\cite{R10}, the approximate formula for the forward--backward
asymmetry obtained in Ref.~\cite{R6} was slightly incorrect.
The result was so close to the correct prediction
that  numerical cross--checks failed to reveal
this tiny discrepancy.
The effect on the asymmetry was of order $10^{-4}$ for $b$
quarks and thus about
two orders of magnitude below the present experimental
error, for $c$ quarks smaller still by another order of magnitude.}
in the mass parameter $\mu$ \cite{R7,R10A},
\begin{eqnarray}
C_F^Q \approx 1- \frac{8}{3} \mu \ + \ \cdots
\end{eqnarray}
In turn, if the thrust axis is chosen as the reference direction, the mass
effects are slightly less pronounced. Note also that it is very
difficult to obtain analytical results for $C_F^Q$ in this case
due to complicated  phase space integrals.

\subsection*{4. Two--jet forward--backward asymmetry}

The two--jet asymmetry is defined for all events in which
the invariant mass of the jets is less than a fraction $\sqrt{y}$ of    the
total energy, i.e.
\ben
(p+\bar{p})^2 / s < y \ &\longrightarrow & \ z > 1-y   \\
(p+p_g)^2 / s < y \ &\longrightarrow & \ \bar{x} >1+
\scriptstyle \frac{1}{4} \displaystyle
\mu^2 -y
\non \\
(\bar{p}+p_g)^2 / s <y \ &\longrightarrow & \ x >1+
\scriptstyle \frac{1}{4} \displaystyle
\mu^2 -y
\non
\een

\nn The two--jet forward--backward asymmetry may then be written again
as
\beq
A_{FB}^Q |_{\rm 2jets} = A_{FB/0}^Q \left[
                                            1 - \frac{\alpha_s}{\pi}
C_F^Q \right]
\eeq
with the coefficients $C_F^Q$ given in the Table 2 and 3 for a sample
of $y$ values\footnote{The entries in Tables 2/3
supersede those given
in Ref.~\cite{R6}.}; the reference axis is chosen
to be the quark axis in Table 2 and
the thrust axis in Table 3. The quark masses
have been defined as in Table 1. The entries to the left of the bars
again correspond to the pole quark masses, those to the right to
the $\overline{MS}$ mass definitions at the scale $M_Z$.  \s

\begin{table}[hbt]
\begin{center}
\begin{tabular}{|c||c|c|c|} \hline \hline
& & & \\
$ \ \ \ \ y_{\rm cut} $ \ \ \ \ & \ \ \ \ $C_F^{m=0} $ \ \ \ \ &   \ \ \ \
$C_F^{c}$ \ \ \ \ & \ \ \ \ $C_F^b$ \ \ \ \ \\
     &       &       &       \\ \hline
0.01 & 0.12 & 0.06 $|$ 0.09 & 0.02 $|$ 0.03 \\
0.02 & 0.21 & 0.14 $|$ 0.17 & 0.05 $|$ 0.08 \\
0.04 & 0.35 & 0.28 $|$ 0.31 & 0.14 $|$ 0.20 \\
0.08 & 0.55 & 0.48 $|$ 0.52 & 0.34 $|$ 0.40 \\
0.16 & 0.81 & 0.74 $|$ 0.77 & 0.60 $|$ 0.66 \\   \hline
2/3  & 1.00 & 0.93 $|$ 0.96 & 0.80 $|$ 0.86 \\   \hline
\end{tabular}
\end{center}
\caption{\it
      Coefficient $C_F^Q$ of $(-\alpha_s/\pi)$ in $A_{FB}^Q$ as a
function of the invariant mass of two jets; the reference axis is taken
to
be the \underline{quark} axis; left: pole masses, right:
$\overline{MS}$ masses at the scales $M_Z$.}
\end{table}

\begin{table}[hbt]
\begin{center}
\begin{tabular}{|c||c|c|c|} \hline
& & & \\
$ \ \ \ \ y_{\rm cut} $ \ \ \ \ & \ \ \ \ $C_F^{m=0} $ \ \ \ \ &   \ \ \ \
$C_F^{c}$ \ \ \ \ & \ \ \ \ $C_F^b$ \ \ \ \ \\
     &       &       &       \\ \hline\hline
0.01 & 0.07 & 0.04 $|$ 0.06 & 0.02 $|$ 0.03 \\
0.02 & 0.13 & 0.10 $|$ 0.11 & 0.04 $|$ 0.06 \\
0.04 & 0.23 & 0.19 $|$ 0.21 & 0.11 $|$ 0.15 \\
0.08 & 0.39 & 0.35 $|$ 0.37 & 0.26 $|$ 0.30 \\
0.16 & 0.64 & 0.60 $|$ 0.62 & 0.51 $|$ 0.55 \\  \hline
2/3  & 0.89 & 0.86 $|$ 0.88 & 0.77 $|$ 0.81 \\  \hline
\end{tabular}
\end{center}
\caption{\it
Coefficient $C_F^Q$
of $(-\alpha_s/\pi)$ in $A_{FB}^Q$ as a function of
the invariant mass of two jets; the reference axis is taken to be the
\underline{thrust} axis; left: pole masses, right: $\overline{MS}$
masses at the scale $M_Z$.}
\end{table}

The following conclusions can be drawn from the tables. First,
                          as anticipated, the two--jet
coefficient $C_F^Q$ decreases with decreasing $y$ cut, and
they become very small for
small invariant masses. Two--jet asymmetries are therefore less affected by
the QCD corrections and they can
                              be used to reduce the errors coming
from the relatively poor knowledge of $\alpha_s$
                       and from higher--order effects.
Furthermore, the coefficient $C_F^Q$ of  the two--jet asymmetry
follows the same pattern
as the asymmetry  for single quark jets and for the thrust axis
                                        in Table 1: The QCD corrections
are smaller in the
case where the thrust axis is chosen as the reference axis,  than
in the case where the reference axis is the quark axis;
                                      the corrections
decrease also with increasing mass.
Finally, for the maximum value of the $y$ cut, $y=2/3$, there are
 by definition no three-jet events left over and the two--jet asymmetry
approaches the single--jet forward--backward  asymmetry. \s

\subsection*{5. Summary}

We have presented an overview of the one--loop QCD corrections
to forward--backward asymmetries for heavy--quark
                                                 production in $Z$ decays.
A complete set of results has been given
                       which improve the theoretical predictions for
these asymmetries, the measurement of which provides some of the most precise
determinations
of the electroweak mixing angle $\sin^2\theta_W$ in $Z$ decays
at LEP and SLC.
We have focussed on the consequences of quark--mass corrections
                        which we have shown
to be rather important. The difference between the
QCD corrections to the forward--backward asymmetries for light quarks and for
$b$-quarks turns out to
               be as large as 20\%. With the high--accuracy of the measurements
performed at LEP, these effects must be taken into account properly.
\s

We have discussed these radiative corrections for the
                               two cases where the
reference directions are taken either
                               to be the quark axis or the thrust axis.
In the second case
the QCD corrections are slighty smaller. We have also considered the
consequences of constraints on the jets by demanding their
invariant mass to be
less than a fraction $\sqrt{y}$
      of the total energy. These two-jet
asymmetries are affected less by QCD corrections than the
single--jet asymmetries, and they could
therefore be exploited to minimize
 the errors due to strong interactions in experimental
                                          high--statistics
 analyses. \s

\bigskip

\nn {\bf Acknowledgement:}
We would like to thank R.~Tenchini for helpful comments. PMZ
thanks D.~Yu.~Bardin and J.B.~Stav for clarifying discussions on
the expansion of the single--jet $FB$ asymmetries in the quark mass,
and M.~Bilenky and A.~Santamaria for a correspondence.

\bigskip

\nn

\newpage

\subsection*{Figure Caption}

Fig.~1 Definition of the oriented thurst axis $\vec{T}$ in 3-jet events.

\begin{figure}[hbt]
\vspace*{0cm}
\hspace*{.3cm}
\epsfxsize=15cm \epsfbox{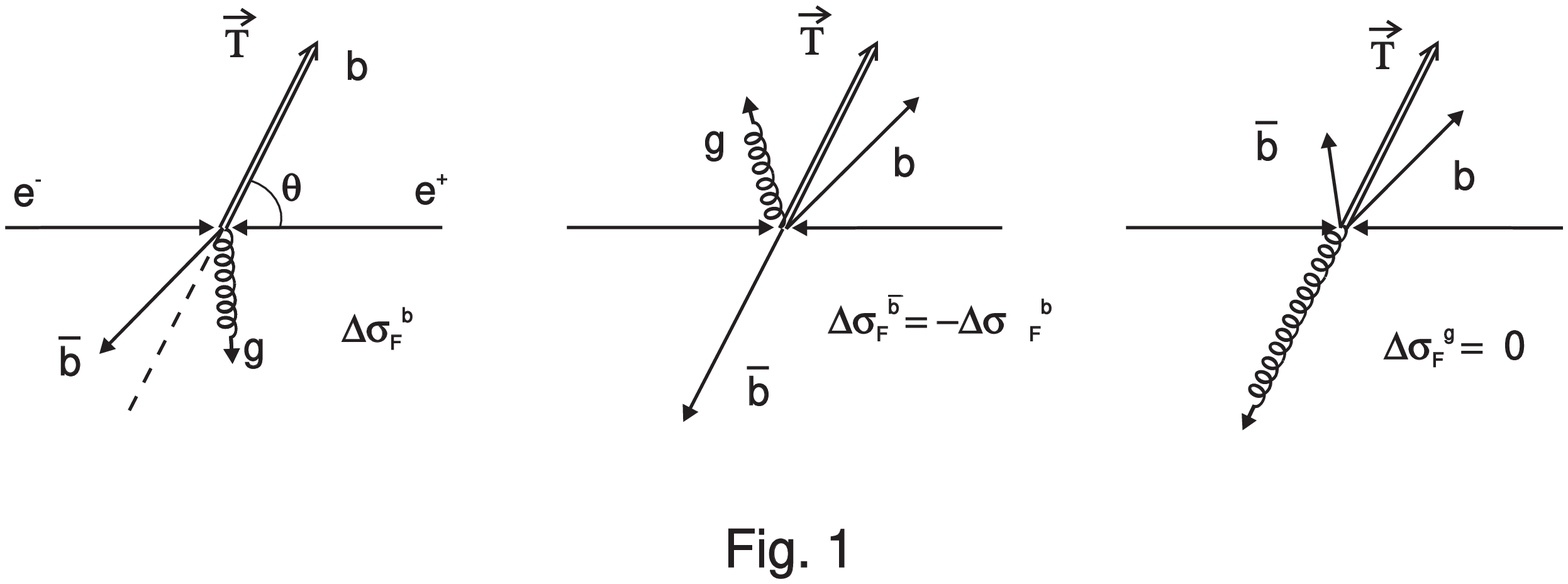}
\vspace*{0cm}
\end{figure}


\vspace{1.5cm}

\begin{thebibliography}{99}

\bibitem{R1} J. H. K\"uhn, P. M. Zerwas
           [{\it conv.}] {\it et al.}, Heavy Flavor Report in the
Proceedings ``Z Physics at LEP",
CERN 89--08 (1989), G. Altarelli {\it et al.} ({\it eds});
A.~Djouadi {\it et al.},
             Proceedings ``High-Luminosity                              at
LEP",
CERN 91--1 (1989), E. Blucher {\it et al.} ({\it eds});
A. Blondel, F.~M.~Renard and C. Verzegnassi, Proceedings
``Polarization at LEP", CERN 88--06 (1988), G. Alexander
{\it et al.} ({\it eds}).

\bibitem{R2} D.~Buskulic {\it et al.} [Aleph],
Phys. Lett. B335 (1994) 99;
P.~Abreu {\it et al.} [Delphi], CERN-PPE/94-161 and Z. Phys. C
{\it in print};
M.~Acciarri {\it et al.} [L3], Phys. Lett. B335 (1994) 544;
R.~Akers {\it et al.} [Opal], Physics Note PN118 (1994).


\bibitem{R3} M.~B\"ohm and W.~Hollik
                  [{\it conv.}] {\it et al.}, Report of the
``Forward-Backward
Asymmetries" Group in the Proceedings of the Workshop ``Z Physics at LEP",
CERN 89--08, G. Altarelli {\it et al.} ({\it eds}).

\bibitem{R3A} E.~Laermann, K.H.~Streng and P.~M.~Zerwas, Z. Phys.
C3 (1980) 289; E {\it ibid} C52 (1991) 352;
J.~Jers\'{a}k, E. Laermann and P.M.~Zerwas, Phys. Lett. B98 (1981) 363.

\bibitem{R4} J.~Jers\'{a}k, E.~Laermann and P.~M.~Zerwas, Phys.~Rev. D25
(1980) 1218.

\bibitem{R5} A.~Djouadi, Z.~Physik C39~(1988)~561.

\bibitem{R6} A.~Djouadi, J.~H.~K\"uhn and P.~M.~Zerwas, Z.~Physik C46
(1990) 411.

\bibitem{R7} A.~B.~Abruzov, D.~Yu.~Bardin and A.~Leike, Mod. Phys. Lett.
A7 (1992) 2029; E $ibid$ A9 (1994) 1515.

\bibitem{R8} G.~Altarelli and B. Lampe, Nucl. Phys. B391 (1993) 3.

\bibitem{R9} B. Lampe, Preprint MPI--Ph/93--74 (1993).

\bibitem{R10} E.~Laermann, Diploma Thesis, RWTH Aachen, 1986.

\bibitem{R10A} J.~Stav, private communication;
see also J.~Stav and H.~Olsen, Preprint Trondheim 1994--17.

\bibitem{R10B} A. Djouadi and C. Verzegnassi, Phys. Lett. B195 (1987)
265;
A. Djouadi, Nuovo Cim. 100A (1988) 357.

\bibitem{R11} D. Buskulic {\it et al.} [Aleph], Z. Phys. C62
(1994) 179.

\bibitem{R11A} K.~G.~Chetyrkin and J.~H.~K\"uhn, Phys. Lett. B248
(1990) 359;
M.~Bilenky, G.~Rodrigo and A.~Santamaria, Preprint CERN--TH.7419/94;
Y.~Koide, Shizuoka Preprint US--94--05.


\end{thebibliography}
\end{document}